\documentclass[showpacs,prb,amsmath,amssymb]{revtex4}% Physical Review B

\usepackage{graphicx}% Include figure files
\begin{document}

%\preprint{APS/123-QED}

\title{Separation Between Antiferromagnetic and Ferromagnetic Transitions in
Ru$_{1-x}$Cu$_{x}$Sr$_{2}$EuCu$_{2}$O$_{8+\delta}$}

\author{Y.~Y.~Xue$^{1}$}
\author{F.~Chen$^{1}$}
\author{J.~Cmaidalka$^{1}$}
\author{R.~L.~Meng$^{1}$}
\author{C.~W.~Chu$^{1,2,3}$}
\affiliation{$^{1}$Department of Physics and Texas Center for Superconductivity, 
University 
of Houston, 202 Houston Science Center, Houston, Texas 77204-5002}
\affiliation{$^{2}$Lawrence Berkeley National Laboratory, 1 Cyclotron Road, 
Berkeley, California 94720}
\affiliation{$^{3}$Hong Kong University of Science and Technology, 
Hong Kong}

\date{\today}% It is always \today, today,

\begin{abstract}
The macroscopic magnetizations of Ru$_{1-x}$Cu$_{x}$Sr$_{2}$EuCu$_{2}$O$_{8+\delta}$ with 
$x$ between 0 and 0.15 were investigated. A ferromagnet-like transition as well as an 
antiferromagnet-like transition appear around $T_{M}$ in the low-field magnetization and 
around $T_{AM}$ in the high-field differential susceptibility, respectively. The separation 
between them, which is accompanied by a flat plateau in the magnetic $C_{p}$, increases with 
$x$. Superparamagnetic $M(H)$ and slow spin dynamics, i.e. characteristics of nanomagnetic 
clusters, were observed far above $T_{M}$. A comparison with 
RuSr$_{2}$(Eu$_{1-y}$Ce$_{y}$)Cu$_{2}$O$_{10+\delta}$ and some manganites further suggests 
that a phase separation occurs, which can describe well the conflicting 
magnetic-superconductivity data previously reported.
\end{abstract}

\pacs{74.81.-g, 74.72.-h, 75.40.Cx}% PACS, the Physics and Astronomy
\maketitle

The puzzling bulk, yet granular, superconductivity (SC) in rutheno-cuprates 
RuSr$_{2}$RCu$_{2}$O$_{8+\delta}$ (Ru1212R) and RuSr$_{2}$(R,Ce)$_{2}$Cu$_{2}$O$_{10+\delta}$ 
(Ru1222R) with R = Gd, Eu, or Y,\cite{fel,ber,tal,xue} which coexists with a weak 
ferromagnetism (FM), is closely related to their magnetic structure. While a homogeneous 
canted antiferromagnetic (CAFM) spin-order may coexist with more-or-less ordinary 
superconductivity, such as the proposed Meissner state or the $\pi$-phase SC,\cite{ber,pic} 
magnetic inhomogeneity at length scales $\ge \xi$ will unavoidably lead to a 
Josephson-junction-array-like superconductivity,\cite{xue} where $\xi$ is the coherence 
length. In the case of Ru1222R, the reported data seem to indicate a rather complicated 
magnetic structure. Both the AFM-like differential-susceptibility maximum of the Ru 
($\chi_{Ru~only}$) and the hyperfine splitting of the M\"{o}ssbauer spectra, for example, 
occur at temperatures almost two times higher than the $T_{M}$, where an FM-like transition 
occurs in the low-field field-cooled magnetization ($M_{FC}$).\cite{fel,xue2} Either a 
phase separation or a multistage transition, therefore, should occur.\cite{fel,xue2} On the 
other hand, the situation of Ru1212R has been suggested to be different. The inflection point 
$T_{AM}$ at $\partial^{2}(T\chi_{Ru~only})/\partial T^{2} = 0$, which should be the N\'{e}el 
temperature in simple antiferromagnets,\cite{fis} and the $T_{M}$ are in rough agreement for 
a Ru1212Eu sample.\cite{but} Mean-field-like scaling has also been observed below $T_{M}$ by 
both neutron powder diffraction (NPD) and zero-field nuclear magnetic resonance 
(ZFNMR).\cite{lyn,tok} It is therefore natural that a simple CAFM was assumed in many 
previous investigations. This model, however, faces a dilemma in accommodating the 
magnetizations and the ZFNMR and NPD data. The NPD, for example, indicated that the Ru spins 
are AFM-aligned (G-type) along the $c$ axis with a very tight upper limit of the FM 
components, i.e. $< 0.1$ and $\approx 0.2$~$\mu_{B}$/Ru at $H = 0$ and 0.4~T~$\le H \le$ 7~T, 
respectively.\cite{lyn} The spontaneous magnetization of the sample, however, reaches 
$M_{r} \approx 800$~emu/mole, i.e. an FM component 0.28~$\mu_{B}$/Ru at $H = 0$. The 
extrapolated zero-field magnetization of 0.6~$\mu_{B}$/Ru at 50~K,\cite{but} which may serve 
as a lower limit for the FM component at 5~T, is again three times larger. The ZFNMR data, 
in addition, demonstrate that the Ru-spins should be aligned perpendicular to the $c$ with a 
major (or dominant) FM component.\cite{tok} This unusual magnetic structure, which appears as 
G-type AFM along $c$ in NPD but ordered along $a,b$ with a large FM component in both 
magnetization and NMR, suggested that the magnetic structure of Ru1212R deserved a 
reexamination. It should be pointed out that both the extremely broad $C_{p}$ peak and the 
superparamagnet-like $M(H)$ up to 2~$T_{M}$ in Ru1212Gd already suggest that its magnetic 
transition is far from simple:\cite{tal,but} the spin correlations may exist up to 2~$T_{M}$ 
with a significant entropy and a correlation size as large as 
$10^{2}$--$10^{3}$~$\mu_{B}$,\cite{xue2} both being characteristic of phase separation. It is 
interesting to note that both $T_{M}$ and $T_{AM}$ of Ru1212R can be tuned by 
Cu-doping.\cite{kla} The evolutions of $M$, $\chi_{Ru~only}$, and $C_{p}$ of 
Ru$_{1-x}$Cu$_{x}$Sr$_{2}$EuCu$_{2}$O$_{8+\delta}$ with $0 \le x \le 0.15$, therefore, were 
measured. The $T_{M}$ drops more than 25~K with $x$ while the variation in $T_{AM}$ is 
negligibly small. A separation between $T_{M}$ and $T_{AM}$ is developed with $x$. This 
separation is further accompanied by a magnetic $C_{p}/T$ with a flat plateau between $T_{M}$ 
and $T_{AM}$. Hence, a mesoscopic phase-separation is suggested.

\begin{figure*}
\includegraphics[scale=.5]{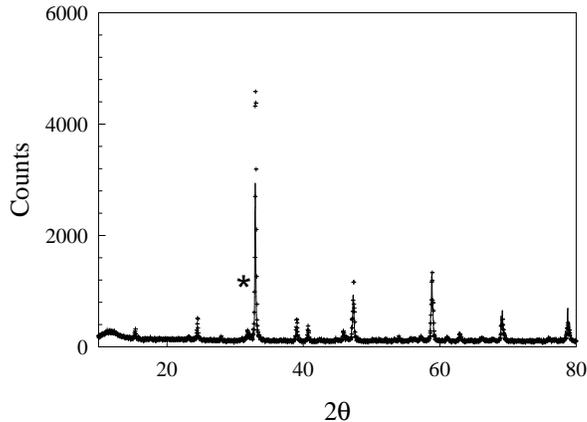}% Here is how to import EPS art
\caption{\label{fig1}The XRD of a (Ru$_{0.9}$Cu$_{0.1}$)Sr$_{2}$EuCu$_{2}$O$_{8+\delta}$ 
sample. $+$: data; solid line: the Rietveld fit; *: the impurity lines.}
\end{figure*}

Ceramic Ru$_{1-x}$Cu$_{x}$Sr$_{2}$EuCu$_{2}$O$_{8+\delta}$ samples with $x$ between 0 and 0.15 
were synthesized following the standard solid-state-reaction procedure. Precursors were first 
prepared by calcinating commercial oxides at 600--900~$^{\circ}$C under flowing O$_{2}$ at 
1~atm. Mixed powder with a proper cation-ratio was then pressed into pellets and sintered at 
960~$^{\circ}$C. The final heat treatment was done at 1065--1070~$^{\circ}$C for 7 d in oxygen 
after repeatedly sintering and regrinding.\cite{xue} The structure of the samples was 
determined by powder X-ray diffraction (XRD) using a Rigaku DMAX-IIIB diffractometer. The 
$x$ dependence of the lattice parameters, i.e. the $c \approx 11.553(2)$ to 11.550(2) {\AA} for 
$x = 0$ and 0.15, respectively, is slightly weaker than that reported for 
Ru$_{1-x}$Cu$_{x}$Sr$_{2}$GdCu$_{2}$O$_{8+\delta}$.\cite{kla} Minor impurity phases, likely 
SrRuO$_{3}$ or oxides of (Sr,Cu), are below 5\% at $x \le 0.15$ (Fig.~\ref{fig1}). The 
composition was measured by a JEOL JXA 8600 electron microprobe with attached wavelength 
dispersive spectrometers (WDS). The local inhomogeneity of $1-x$ is within the experimental 
resolution of $\pm 0.05$.\cite{lor} The magnetizations were measured using a Quantum Design 
SQUID magnetometer with an $ac$ attachment and the specific heat was measured in a 
Quantum Design PPMS with a specific-heat attachment.

\begin{figure*}
\includegraphics[scale=.5]{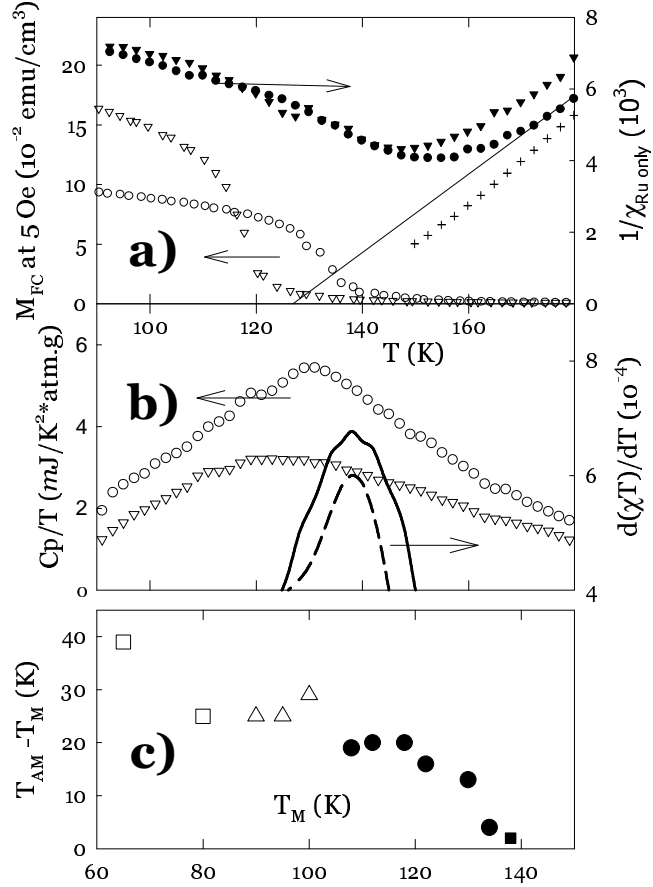}% Here is how to import EPS art
\caption{\label{fig2}a) The magnetizations. For the $x = 0$ sample, $\circ$: $M_{FC}$(5~Oe); 
$+$: $H/M_{FC}$(1~T); $\bullet$: $1/\chi_{Ru~only}$; solid line: C-W fit. For the $x = 0.1$ 
sample, $\bigtriangledown$: $M_{FC}$(5~Oe); $\blacktriangledown$: $1/\chi_{Ru~only}$. b) The 
spin entropy. $\circ$: the magnetic $C_{p}/T$ of the $x = 0$ sample; $\bigtriangledown$: that 
of the $x = 0.1$ sample; solid line: $\partial(T\chi)/\partial T$ of the $x = 0$ sample; 
dashed 
line: that of the $x = 0.1$ sample. c) The evolution of $T_{AM}-T_{M}$ with $T_{M}$. 
$\bullet$: Cu-doped Ru1212Eu; $\blacksquare$: Ru1212Eu of Butera \textit{et al.};\cite{but} 
$\bigtriangleup$: annealed Ru1222Gd; $\square$: as-synthesized Ru1222Eu.}
\end{figure*}

Superconductivity appears in all the samples below a critical temperature $T_{c} \approx$ 
20--30~K. A single-step jump of $M_{FC}$ also appears with cooling at a higher temperature 
(Fig.~\ref{fig2}a). According to the scaling correlation 
$(HM_{0}/MH_{0})^{1/\gamma} = t + (M/M_{0})^{1/\beta}$, the $\partial M/\partial T$ of an 
ideal ferromagnet should decrease with $t = (T-T_{M})/T_{M}$ as $1/t^{\gamma+1}$ above 
$T_{M}$, but increase as $(-t)^{1-\beta}$ below, where $0 < \beta < 1$, $\gamma > 0$, $H_{0}$, 
and $M_{0}$ are two critical exponents and two critical amplitudes, respectively. The 
situation for a CAFM magnet should be similar. Therefore, the inflection point of $M_{FC}(T)$ 
at 5~Oe, i.e. the temperature at which $\partial M_{FC}/\partial T$ peaks, is used as the 
$T_{M}$ (Fig.~\ref{fig2}a). The well defined $T_{M}$ and the large FM component below $T_{M}$ 
are in rough agreement with those reported for Ru1212Eu,\cite{but} but rather different from 
those of Ru$_{1-x}$Cu$_{x}$Sr$_{2}$GdCu$_{2}$O$_{8+\delta}$,\cite{kla} where no clear FM 
transition can be identified with $x \ge 0.1$. Differences in both the rare-earth elements and 
the synthesis procedures may contribute to the variation. It should be pointed out that the 
well defined $T_{M}$ and the large $M_{FC}$ of our samples make the analysis of $M_{FC}$ and 
$C_{p}$ easier and without significant interference from the minor impurities. A systematic 
decrease of $T_{M}$ with $x$ is observed, e.g. $T_{M} \approx 134$~K and 117~K at $x = 0$ and 
0.1, respectively (Fig.~\ref{fig2}a). It is also interesting to note that the reported 
bifurcation point between $M_{ZFC}$ and $M_{FC}$, which should be very close to $T_{M}$ if 
the domain pinning is strong, in Ru$_{1-x}$Cu$_{x}$Sr$_{2}$GdCu$_{2}$O$_{8+\delta}$ shows 
almost the same $x$ dependence, i.e. down to $\approx 115$~K and 100~K with $x = 0.1$ and 0.2, 
respectively.\cite{kla}

The $\chi = M/H$ of a simple AFM magnet, which will be $H$-independent far above its AFM 
transition, should have a maximum slightly above the N\'{e}el temperature, $T_{AM}$. It has 
been suggested, in fact, that the magnetic energy, $E_{m}$, and $\chi$ should both depend on 
the pair correlation functions $\Gamma(r) = 3[S^{z}(0)S^{z}(r)]/S(S+1)$ as 
$E_{m} \propto \Gamma_{1}$ and 
$\chi \propto [1+ \Sigma_{r} \Gamma(r)]/T \approx [1+f(T)\Gamma_{1}]/T$, where $\Gamma(r)$, 
$\Gamma_{1}$, and $f(T)$ are the pair-correlation with the pair distance $= r$, the 
correlation with the nearest-neighbor, and a slowly varying function of $T$, 
respectively.\cite{fis} This leads to an approximation of 
$C_{p} \propto \partial(T\chi)/\partial T$ if the short-range correlation $\Gamma_{1}$ 
is dominant. $T_{AM}$, therefore, can be defined as the temperature of the 
$\partial(T\chi)/\partial T$ peak,\cite{fis} which is observed close to the $\chi$-maximum 
temperature in 3D but much lower in 2D.\cite{jon} For CAFM magnets, an FM-like $M_{FC}$ step 
may coexist with a $\partial(T\chi)/\partial T$ peak. However, $T_{AM} \approx T_{M}$ is 
expected, except for the possible $H$-induced transition shifts.\cite{note} 

To analyze the magnetization of Ru$_{1-x}$Cu$_{x}$Sr$_{2}$EuCu$_{2}$O$_{8+\delta}$, the 
Eu/CuO$_{2}$ contributions were first eliminated using the procedure previously 
proposed,\cite{but} i.e. with a Van Vleck susceptibility of free Eu$^{3+}$ and a 
$T$-independent $\chi_{0}$ of $8.7 \times 10^{-4}$~emu/mole for CuO$_{2}$. For the undoped 
sample with $x = 0$, the Ru contribution is $H$-independent and follows a Curie-Weiss (C-W) 
fit only above 250~K with a C-W constant $\approx 2.6$~$\mu_{B}$/Ru and a Curie temperature of 
127~K. Deviation from the C-W fit and large superparamagnetic $M(H)$, however, develop at 
lower temperatures (Fig.~\ref{fig2}a). The Ru contribution to $\chi$ at 1~T, for example, is 
more than 10\% higher than that expected between 180~K and $T_{AM}$ (Fig.~\ref{fig2}a), 
indicating a dominant FM interaction. The 5~T differential Ru-susceptibility after subtracting 
the Eu/CuO$_{2}$ contributions ($\chi_{Ru~only}$), however, shows an opposite downturn, 
suggesting significant AFM interactions (Fig.~\ref{fig2}a). In particular, a minimum of 
$1/\chi_{Ru~only}$ and a $\partial(T\chi)/\partial T$ peak appear around 157~K 
(Fig.~\ref{fig2}a) and $T_{AM} \approx 138$~K (Fig.~\ref{fig2}b), respectively, for the 
$x = 0$ sample. Undoped Ru1212Eu, therefore, might be interpreted as a simple CAFM by 
either ignoring the 4~K difference between $T_{M}$ and $T_{AM}$,\cite{but} or by regarding it 
as a small $H$-induced transition shift. 

To further confirm the presumed $T_{AM}$, the magnetic specific heat was measured at a zero 
field using a nonsuperconducting YBa$_{2}$(Cu$_{2.73}$Zn$_{0.27}$)O$_{7}$ (YBCO) ceramic as 
the reference (Fig.~\ref{fig2}b). The raw specific heat of Ru1212 is well above that of YBCO 
between 80 and 180~K, but the two merge outside this region, a situation similar to the data 
of Ru1212Gd.\cite{tal} The magnetic $C_{p}/T$, i.e. the difference between Ru1212R and YBCO, 
shows a well defined peak at 133~K, which is only slightly lower than the 138~K 
$\partial(T\chi)/\partial T$ peak observed. This agreement between the $C_{p}/T$ peak at 
zero field and the $\partial(T\chi)/\partial T$ peak at 5~T again demonstrates that the 
procedure of Fisher\cite{fis} works reasonably well and that the $H$-induced transition shift 
is small in our case. It is also interesting to note the high-$T$ tail of $C_{p}/T$ and 
the non-C-W magnetization up to 180~K or higher (Figs.~\ref{fig2}a,b). Significant short-range 
spin orders, therefore, should occur far above $T_{M}$ and $T_{AM}$.

With the Cu doping, however, the $T_{M}$ and the $T_{AM}$ evolve in different ways and the 
separation between them broadens. At $x = 0.1$, for example, the $T_{M}$ is quickly suppressed 
to 117~K but the $\partial(T\chi)/\partial T$ peak remains at 138~K (Figs.~\ref{fig2}a,b). 
The accompanying $C_{p}/T$ appears to broaden with $x$ as well (Fig.~\ref{fig2}b). In 
particular, the well defined peak evolves into a flat plateau between $T_{M}$ and $T_{AM}$ 
(Fig.~\ref{fig2}b). It should also be pointed out that the separation at $x = 0.1$ is larger 
than the transition width in $M_{FC}$. Neither the sample inhomogeneity nor the experimental 
resolution, therefore, can account for the separation (Figs.~\ref{fig2}a,b). The AFM-like 
$\partial(T\chi)/\partial T$ peak and the FM-like $M_{FC}$ jump seem to carry 
distinct spin entropies of comparable strength.

It is therefore interesting to compare the data with that of Ru1222R, where two separate 
transitions have been observed in both magnetizations and M\"{o}ssbauer 
spectra.\cite{fel,xue2} The $T_{M}$ and $T_{AM}$ of 
Ru$_{1-x}$Cu$_{x}$Sr$_{2}$EuCu$_{2}$O$_{8+\delta}$ samples with $0 \le x \le 0.15$, the 
O$_{2}$/Ar-annealed RuSr$_{2}$(Gd$_{1.4}$Ce$_{0.6}$)Cu$_{2}$O$_{10+\delta}$, and two 
as-synthesized Ru1222Eu samples are shown in Fig.~\ref{fig2}c.\cite{xue2} The separation 
$T_{AM}-T_{M}$ increases systematically with decreasing $T_{M}$ in the Cu-doped Ru1212Eu: from 
an extrapolated zero-separation at $T_{M} \approx 140$~K to 25~K at $T_{M} \approx 110$~K, 
where the data smoothly evolve into that of Ru1222R (Fig.~\ref{fig2}c). The observation of 
$T_{AM} = T_{M}$ in the Ru1212Eu sample,\cite{but} therefore, may be only a coincidence. 
Distinct AFM and FM transitions may coexist in both Ru1212R and Ru1222R. 

These two transitions, as has been argued in the case of Ru1222R,\cite{fel,xue2} may be due to 
either a mesoscopic phase-separation or a multistage transition. The magnetic properties 
between $T_{M}$ and $T_{AM}$, however, will be different in these two scenarios: some parts 
of Ru1212R 
should be in superparamagnetic states during phase separation, but should stay in a long-range 
spin-order state during a multistage transition. Evidence for the possible phase-separation in 
Ru1222R, for example, is found in both the superparamagnetic $M(H)$ with a magnetic 
cluster-size of $10^3$~$\mu$B and the slow spin dynamics far above $T_{FM}$.\cite{xue2} 
Similar properties were therefore tested in the Cu-doped Ru1212Eu.

\begin{figure*}
\includegraphics[scale=.5]{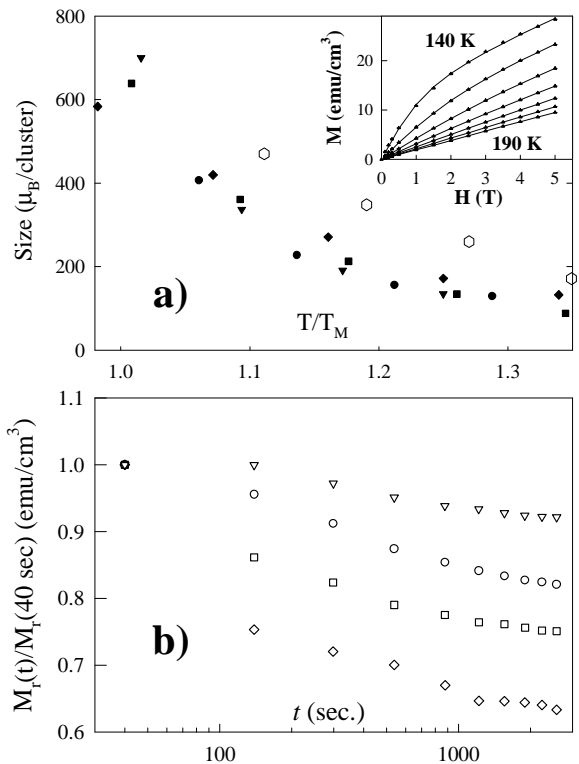}% Here is how to import EPS art
\caption{\label{fig3}a) The cluster sizes for samples with $\bullet$: $x = 0$; 
$\blacktriangledown$: $x = 0.05$; $\blacksquare$: x=0.10; and $\blacklozenge$: x=0.15 and for 
$\circ$: as-synthesized Ru1222Eu ($\times 4$). Inset: The isothermal $M(H)$ of the $x = 0$ sample. 
b) The relaxation of the remnant magnetizations at 160, 150, 140, and 130~K (from top to 
bottom) after field-cooling at 50~Oe.}
\end{figure*}

The Langevin function with an additional linear term, 
$a \cdot H+m \cdot [c\tanh(\mu H/k_{B}T)-k_{B}T/\mu H]$, was used to fit the average 
magnetization in a $M$-$H$ loop (inset, Fig.~\ref{fig3}a).\cite{xue2} The fit is reasonably 
good with the deduced $\mu$ between 100 and 700~$\mu_{B}$/cluster (closed symbols in 
Fig.~\ref{fig3}a), which is 4--5 times smaller than those deduced in Ru1222R, but still far 
larger than that expected based on the spin-fluctuations. A cluster of 
400~$\mu_{B} \approx 200$~Ru ions, for example, would be 4--5~nm or larger in an RuO layer. It 
should be further noted that the $\mu$ so-deduced may be only a lower limit of the actual 
cluster/spin-correlation length.\cite{all} The existence of such large clusters at 
$T/T_{M} > 1.1$ will be difficult to be interpreted as a simple fluctuation. This deduced 
size, on the other hand, appears to be too small for a 
crystalline magnet, as is suggested in the multistage transition model. 

The dynamic spin-response was also studied. The logarithmic increase of $M_{ZFC}$ at 5~Oe with 
time is almost unobservable, with the deduced rate of $dlnM/dlnt < 10^{-3}$ well within our 
experimental resolution, where 60~s~$< t <$ 3600~s is the time after the field switch. This is 
rather different from that of Ru1222Eu,\cite{xue2} but in agreement with the unobservable 
relaxation of Ru1212R $ac$-susceptibility reported between 1~s and 100~s.\cite{ziv} The lack 
of relaxations under the above conditions is apparently related to the 
cluster size in Ru1212R (Fig.~\ref{fig3}a), which is 4--5 times smaller and leads to a quicker 
equilibrium. The slow spin dynamics, therefore, should either be explored in a shorter time 
window or after an enhancement of the energy barriers. Several different experimental 
conditions were then tested, and significant nonlogarithmic relaxations were observed in the 
remnant magnetization after a 50~Oe field-cooling (Fig.~\ref{fig3}b). It is interesting to 
note that the energy barriers are $\approx KV_{c}-\mu H$ and $KV_{c}$, respectively, for the 
$M_{ZFC}$ and the remnant magnetization, where $K$ and $V_{c}$ are the magnetic anisotropy 
and the coherent volume, respectively. This may make the remnant magnetization a more 
favorable candidate for investigating the slow dynamics. The strong $T$ dependence of the 
relaxation observed (Fig.~\ref{fig3}b) suggests, in our opinion, that the relaxation observed 
is unlikely an artifact of the SQUID magnetometer, but supports the existence of 
superparamagnetic clusters.

As pointed out earlier, the phase-separation model may also offer a consistent interpretation 
for the conflicting NPD/NMR and superconductivity data reported previously.\cite{xue2} The 
conflict between the NPD and NMR data for the magnetic structure, for example, may be 
attributed to the fact that the two probes have different sensitivities to various magnetic 
species, like those well documented in manganites.\cite{kap} Similarly, the spatial 
separation between AFM and FM species offers a natural mechanism for the unusual 
superconductivity observed.\cite{xue} Superconductivity can coexist with the AFM matrix. The 
finely dispersed FM clusters, on the other hand, depress the local SC order parameter and 
serve as tunnel barriers for the Cooper pairs. The superconductivity, therefore, may retain a 
significant part of the condensation energy, but appears only as a Josephson-junction array. 
Similarly, the critical temperature observed in the transport will naturally be much lower 
than 
that associated with the corresponding $C_{p}$ anomaly,\cite{ber,tal} and can be easily 
suppressed by external fields.\cite{lor} The intragrain penetration depth will also be much 
longer than those expected based on the proposed universal $1/\lambda_{2}(T_{c})$.\cite{xue}

In summary, a systematic separation between $T_{M}$ and $T_{AM}$ is observed in Ru1212Eu with 
Cu-doping, suggesting the coexistence of FM and AFM orders and the occurrence of a 
mesoscopic phase-separation in the compound. The superparamagnetic $M(H)$ as well as the slow 
spin-dynamics further support the interpretation.

\begin{acknowledgments}
The work in Houston is supported in part by 
NSF Grant No. DMR-9804325, the T.~L.~L. Temple Foundation, the John J. and Rebecca 
Moores Endowment, and the State of Texas through the Texas Center for 
Superconductivity at the University of Houston; and at Lawrence Berkeley 
Laboratory by the Director, Office of Science, Office of Basic Energy Sciences, 
Division of Materials Sciences and Engineering of the U.S. Department of Energy 
under Contract No. DE-AC03-76SF00098.
\end{acknowledgments}


\begin{thebibliography}{99}
\bibitem{fel} I. Felner, U. Asaf, Y. Levi and O. Millo, Phys. Rev. B {\bf 55}, 3374 (1997); 
I. Felner, U. Asaf and E. Galstyan, cond-mat/0111217 (2001).

\bibitem{ber} C. Bernhard, J. L. Tallon, E. Br\"{u}cher and R. K. Kremer, Phys. Rev. B 
{\bf 61}, 14960 (2000).

\bibitem{tal} J. L. Tallon, J. W. Loram, G. V. M. Williams and C. Bernhard, Phys. Rev. B 
{\bf 61}, 6471 (2000).

\bibitem{xue} Y. Y. Xue, B. Lorenz, R. L. Meng, A. Baikalov and C. W. Chu, Physica C 
{\bf 364-365}, 251 (2001); Y. Y. Xue, B. Lorenz, A. Baikalov, D. H. Cao, Z. G. Li and 
C. W. Chu, Phys. Rev. B {\bf 66}, 014503 (2002).

\bibitem{pic} W. E. Pickett, R. Weht and A. B. Shick, Phys. Rev. Lett. {\bf 83}, 3713 (1999).

\bibitem{xue2} Y. Y. Xue, D. H. Cao, B. Lorenz and C. W. Chu, Phys. Rev. B {\bf 65}, 020511 
(2001); Y. Y. Xue, B. Lorenz, D. H. Cao and C. W. Chu, cond-mat/0211342 (2002).

\bibitem{fis} M. E. Fisher, Philos. Mag. {\bf 7}, 1731 (1962).

\bibitem{but} A. Butera, A. Fainstein, E. Winkler and J. Tallon, Phys. Rev. B {\bf 63}, 054442 
(2001).

\bibitem{lyn} J. W. Lynn, B. Keimer, C. Ulrich, C. Bernhard and J. L. Tallon, Phys. Rev. B 
{\bf 61}, 14964 (2000).

\bibitem{tok} Y. Tokunaga, H. Kotegawa, K. Ishida, Y. Kitaoka, H. Takagiwa and J. Akimitsu, 
Phys. Rev. Lett. {\bf 86}, 5767 (2001).

\bibitem{kla} P. W. Klamut, B. Dabrowski, S. Kolesnik, M. Maxwell and J. Mais, Phys. Rev. B 
{\bf 63}, 224512 (2001).

\bibitem{lor} B. Lorenz, Y. Y. Xue, R. L. Meng and C. W. Chu, Phys. Rev. B {\bf 65}, 174503 
(2002).

\bibitem{jon} L. J. de Jongh and A. R. Miedema, Adv. Phys. {\bf 23}, 1 (1974).

\bibitem{note} This is true below a critical field $H_{c}$ since $\partial(T\chi)/\partial T$ 
and $\partial M_{FC}/\partial T$ have similar $T$-dependences. Above $H_{c}$, however, the 
field may break the AFM spin-correlations and a magnetization jump appears. A. Herweijer 
\textit{et al.} reported in Phys. Rev. B {\bf 5}, 4618 (1972), for example, that the 
magnetizations of CsCoCl$_{3}$~$\cdot$~2H$_{2}$O (a 1D CAFM) are linear in $H$ below 0.3~T 
with identical $T_{AM}$ and $T_{M}$.

\bibitem{all} P. Allia, M. Coisson, P. Tiberto, F. Vinai, M. Knobel, M. A. Novak and 
W. C. Nunes, Phys. Rev. B {\bf 64}, 144420 (2001), and references therein.


\bibitem{ziv} I. \v{Z}ivkovi\'{c}, Y. Hirai, B. H. Frazer, M. Prester, D. Drobac, D. Ariosa, 
H. Berger, D. Pavuna, G. Margaritondo, I. Felner and M. Onellion, Phys. Rev. B {\bf 65}, 
144420 (2002).


\bibitem{kap} Cz. Kapusta, P. C. Riedi, M. Sikora and M. R. Ibarra, Phys. Rev. Lett. {\bf 84}, 
4216 (2000).

\end{thebibliography}
\end{document}